\documentclass[pdflatex,sn-nature]{sn-jnl}

\usepackage{graphicx}%
\usepackage{pdfpages}
\usepackage{multirow}%
\usepackage{amsmath,amssymb,amsfonts}%
\usepackage{amsthm}%
\usepackage{mathrsfs}%
\usepackage[title]{appendix}%
\usepackage{xcolor}%
\usepackage{textcomp}%
\usepackage{manyfoot}%
\usepackage{booktabs}%
\usepackage{algorithm}%
\usepackage{algorithmicx}%
\usepackage{algpseudocode}%
\usepackage{listings}%

\usepackage{geometry}
\theoremstyle{thmstyleone}%
\theoremstyle{thmstyletwo}%

\theoremstyle{thmstylethree}%

\begin{document}

\title[Article Title]{Site-orbit masked twin-branch Transformer for symmetry-constrained crystal generation}


\author[1]{Junchi Liu}

\author[1]{Haoran Si}

\author[3]{Luo Yan}

\author[4]{Sergei Tretiak}

\author*[1,2]{Liujiang Zhou}\email{ljzhou@uestc.edu.cn}

\affil[1]{School of Physics, State Key Laboratory of Electronic Thin Films and Integrated Devices, University of Electronic Science and Technology of China, Chengdu 611731, China}

\affil[2]{Institute of Fundamental and Frontier Sciences, University of Electronic Sciences and Technology of China, Chengdu, 611731, China}

\affil[3]{School of mathematics and Physics, University of South China, Hengyang, 421001, China}

\affil[4]{Theoretical Division, Center for Nonlinear Studies, and Center for Integrated Nanotechnologies, Los Alamos National Laboratory, Los Alamos, New Mexico, 87545, United States}

\abstract{
    Generative models for crystal structure prediction (CSP) have expanded the search space for novel materials and improved screening efficiency. However, simultaneously generating stable and novel structures remains challenging because of the complex structural and physical constraints governing crystals. To address this challenge, we propose site-orbit masked twin-branch Transformer model (ST2M), a symmetry-guided discrete--continuous diffusion framework incorporating degree-of-freedom masks and conditioned on Wyckoff patterns. ST2M represents crystal structures as sequences of independent occupied Wyckoff orbits and employs a twin-branch Transformer to jointly denoise discrete and continuous variables. In our experiments, ST2M achieves a structural validity of 96.37\%. For known patterns in dataset, 23.86\% of generated structures simultaneously satisfy stability, uniqueness, and novelty criteria, representing a relative improvement of 68.15\% over the best-performing baseline. First-principles calculations identify a total of 21 previously unreported stable structures among selected candidates with low predicted energies. These results demonstrate that ST2M can generate chemically viable crystal candidates under prescribed crystallographic constraints and generalize to unknown Wyckoff patterns, providing a practical route to accelerated materials discovery.
}

\maketitle
\section*{Introduction}

Discovering functional materials with improved properties is a central objective of materials science, yet the vast chemical and structural search space and the complex relationships between crystal structures and physical properties make this task challenging\cite{bib1,bib2,bib3,bib4,bib5,bib6,bib7,bib8,bib9,bib10,bib11}. Advances in machine learning (ML) have enabled generative approaches to crystal structure prediction (CSP), extending the exploration of candidate materials beyond elemental substitution in known structures~\cite{bib12,bib13}. By learning chemical and geometric regularities from crystal datasets, generative models can efficiently propose candidate compositions and atomic arrangements for subsequent evaluation\cite{bib14,bib15}. These candidates can be incorporated into high-throughput screening workflows, in which ML interatomic potentials provide rapid structural relaxation and energy estimates, followed by first principle calculations to assess the stability and target properties of selected structures~\cite{bib16,bib17}. The practical value of a crystal generator therefore depends on its ability to produce physically valid and stable candidates that are also unique and novel, thereby reducing redundant calculations and improving the efficiency of materials exploration~\cite{bib18,bib19,bib20,bib21}. Existing approaches have made substantial progress in generating plausible crystals, with computational validation identifying stable candidate structures\cite{bib22}. Nevertheless, improving the joint yield of stable, unique, and novel materials remains an important challenge in translating generative performance into effective materials discovery\cite{bib23}.

Crystal generative modeling has progressed from learning structural distributions towards incorporating crystallographic constraints and targeting material properties\cite{bib24,bib25,bib26,bib27,bib28}. crystal diffusion variational autoencoder (CDVAE) established a latent generative framework with a diffusion decoder\cite{bib29}, while DiffCSP advanced joint diffusion of lattices and atomic coordinates\cite{bib30}. MatterGen further extended diffusion models to composition and property conditioned discovery\cite{bib17}. In parallel, the physics guided crystal generative Model (PGCGM) incorporated symmetry transformations and geometric constraints, although atomic overlaps still require subsequent merging\cite{bib12}. Recent symmetry-aware approaches include SymmCD and WyckoffDiff. SymmCD generates asymmetric units using site-symmetry representations, reducing structural redundancy and enabling information across space groups to be shared\cite{bib31}. However, independently assigning compatible positions may require additional checks to avoid coincident occupation of fixed sites. WyckoffDiff, by contrast,  directly generates elemental occupations of Wyckoff positions, enforcing single occupancy of positions with no positional degrees of freedom while leaving the determination of the lattice and coordinate   to a separate step\cite{bib32}. Although these models based on the asymmetric unit representations improve structural control, the coordinate components of their outputs may not strictly align with the anticipated assigned Wyckoff sites and require projection onto compatible positions to satisfy the corresponding site symmetry constraints. Integrating symmetry constraints with complete crystal generation and a high yield of stable, unique, and novel candidates remains an open challenge.

In this work, we propose the site-orbit masked twin-branch Transformer model (ST2M), a crystal generative framework that enforces compatibility with prescribed crystallographic assignments and single occupancy of fixed Wyckoff positions. Generation begins by sampling a space group and a compatible Wyckoff pattern, defined as a sequence of occupied independent crystallographic orbits representing the asymmetric unit. A twin-branch Transformer serves as the denoising network, jointly recovering the discrete elemental identities and continuous structural variables conditioned on this pattern. Coordinate and lattice masks restrict denoising to symmetry allowed free parameters, preserving the prescribed positional relations and lattice constraints throughout generation. The complete crystal is subsequently reconstructed by applying the space group symmetry operations to each orbit representative. Wyckoff patterns can be drawn from the target dataset to generate structures within familiar crystallographic configurations, or sampled according to orbit occupancy frequencies to explore previously unknown configurations while respecting occupancy constraints. We evaluate ST2M through generation benchmarks, large scale screening with a machine learning interatomic potential, and density functional theory (DFT) validation of selected candidates. The results demonstrate its ability to generate valid, stable, and novel structures under both known and unknown patterns, supporting its practical use in materials discovery under specified crystallographic constraints.

\section*{Results}
\subsection*{ST2M framework for symmetry-constrained crystal generation}

Asymmetric unit representations reduce structural redundancy by retaining only crystallographically independent atoms. Some symmetry aware diffusion models, exemplified by SymmCD, jointly generate atomic species, representative coordinates, lattice parameters, and local site symmetries through a common denoising network~\cite{bib31}. Although this design captures correlations across the crystal, it also requires the network to reconcile categorical and continuous variables with different noise statistics and crystallographic constraints, which can make denoising more demanding. Moreover, predicting local site symmetries without prescribing the full occupied Wyckoff pattern does not automatically enforce compatibility among site assignments. These considerations motivate separating the denoising objectives while retaining chemical--geometric coupling and explicit control over site occupancy.

\begin{figure}[t]
    \centering
    \includegraphics[width=\textwidth]{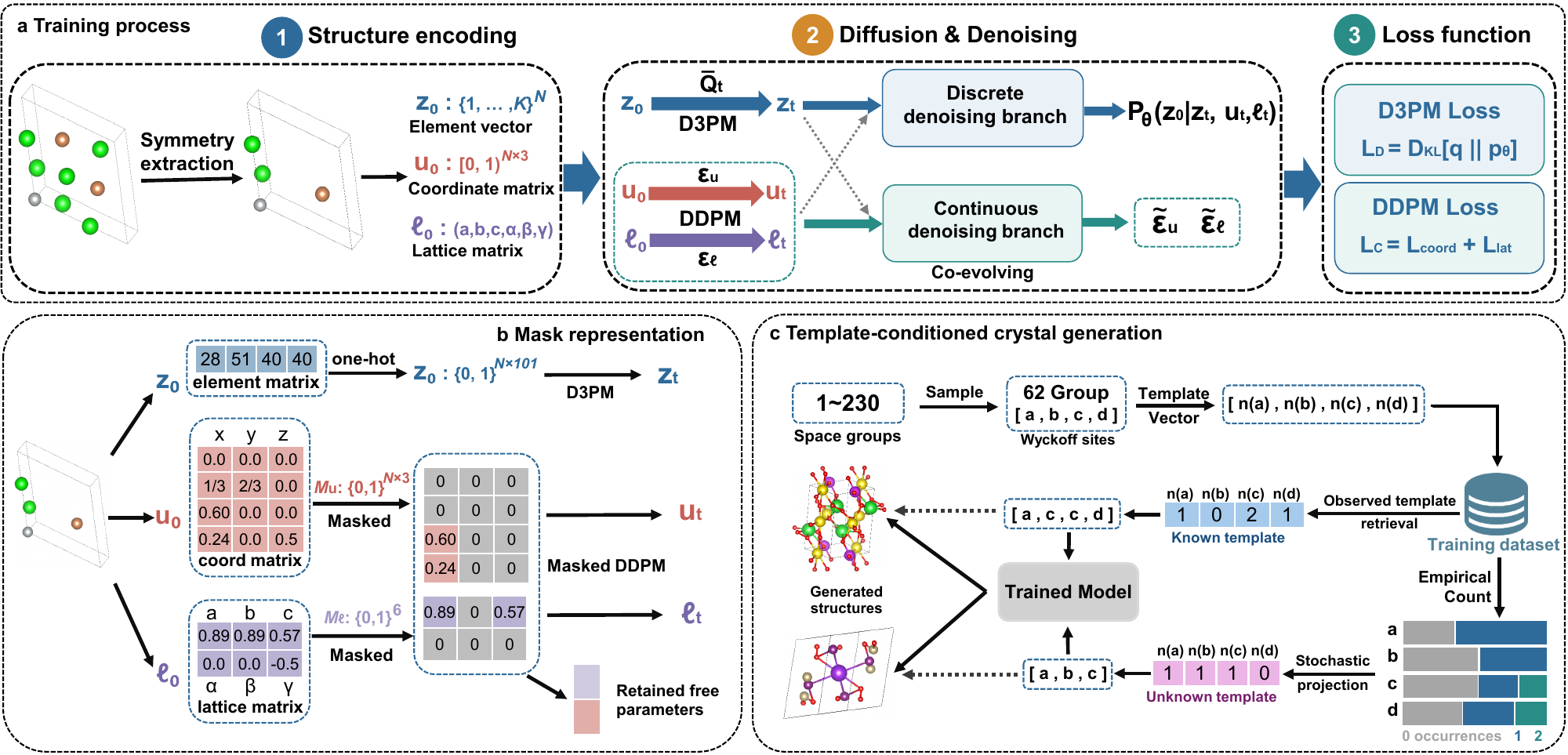}
    \caption{\textbf{Overview of ST2M.} \textbf{a}, Symmetry analysis reduces a crystal to its occupied Wyckoff orbits, including element labels $\mathbf{Z}_0$, representative coordinates $\mathbf{U}_0$, and lattice variables $\boldsymbol{\ell}_0$. Coupled categorical and Gaussian diffusion processes are reversed by discrete and continuous Transformer branches. \textbf{b}, Wyckoff coordinate and lattice masks restrict continuous diffusion to symmetry allowed free variables; constrained components are reconstructed analytically. Element identities for all occupied orbits participate in categorical diffusion. \textbf{c}, Wyckoff pattern conditioned generation uses known Wyckoff patterns retrieved from the training set or unknown Wyckoff patterns sampled from empirical per-position orbit count distributions and screened for novelty. These Wyckoff patterns condition the trained model to generate chemical and structural realizations, followed by symmetry reconstruction and expansion.}
    \label{fig:fig1}
\end{figure}

To address these challenges, we develop ST2M, a symmetry-constrained discrete--continuous diffusion framework with a site-orbit masked twin-branch Transformer. Its two branches specialize in categorical and continuous denoising while sharing the noisy chemical and geometric state, preserving the mutual dependence of composition and structure~\cite{bib33,bib34}. An occupied Wyckoff pattern specifies the crystallographic orbits before diffusion and enforces single occupancy of fixed positions. Figure~\ref{fig:fig1}a outlines the resulting workflow: symmetry-based encoding, coupled diffusion and denoising, and reconstruction of the full crystal. Each standardized input is reduced to one representative atom per independent orbit; for $N$ occupied orbits, the model state is
\begin{equation}
    \mathcal{C}_0=\left(G,\mathcal{W},\mathbf{Z}_0,\mathbf{U}_0,\boldsymbol{\ell}_0\right),
    \label{eq:results_crystal_representation}
\end{equation}
where $G$ specifies the space group and $\mathcal{W}=(w_1,\ldots,w_N)$ is the ordered sequence of occupied Wyckoff positions. Symmetry-equivalent atoms are recovered by applying space group operations after sampling.

Reducing the crystal to independent orbits does not, by itself, eliminate redundancy within the continuous representation. Once the space group and Wyckoff positions are specified, some coordinate components are fixed and others obey exact positional relations; lattice lengths and angles are likewise subject to crystal system constraints. Treating all components as independent denoising targets therefore asks the network to recover information already determined by symmetry, adding unnecessary prediction tasks and competing with the learning of free variables. ST2M removes these redundant targets through the coordinate and lattice masks in Figure~\ref{fig:fig1}b. For each $(G,w_i)$ pair, $\mathbf{M}^{u}_i$ selects only the independent positional parameters: a site of the form $(x,1/2,1/4)$ contributes $x$ alone, whereas a fixed site contributes none. The lattice mask $\mathbf{M}^{\ell}_G$ performs the analogous selection; for a tetragonal cell, for example, only two independent lengths are learned. The masked forward states are
\begin{equation}
    \begin{aligned}
        \mathbf{U}_t&=\mathbf{M}^{u}\odot\left(
        \sqrt{\bar{\alpha}^{C}_t}\,\mathbf{U}_0
        +\sqrt{1-\bar{\alpha}^{C}_t}\,\boldsymbol{\epsilon}^{u}
        \right),\\
        \boldsymbol{\ell}_t&=\mathbf{M}^{\ell}_G\odot\left(
        \sqrt{\bar{\alpha}^{C}_t}\,\boldsymbol{\ell}_0
        +\sqrt{1-\bar{\alpha}^{C}_t}\,\boldsymbol{\epsilon}^{\ell}
        \right).
    \end{aligned}
    \label{eq:results_masked_diffusion}
\end{equation}
where $\bar{\alpha}^{C}_t$ is the cumulative continuous noise schedule, $\odot$ denotes element-wise multiplication, $\mathbf{M}^{u}$ collects the orbit-level coordinate masks, and $\boldsymbol{\epsilon}^{u}$ and $\boldsymbol{\epsilon}^{\ell}$ are standard Gaussian noise. The same masks restrict reverse updates and noise-prediction losses to the independent components. Inactive entries remain zero during diffusion, and their physical values are restored analytically from the Wyckoff relations and lattice constraints. The network thus learns the structural freedom left by symmetry without repeatedly estimating its predetermined components.

The usefulness of this pattern conditioned formulation also depends on obtaining orbit combinations that are plausible yet extend beyond arrangements already observed in training. Unrestricted recombination can yield poorly supported occupancy patterns, whereas using observed patterns alone limits exploration. We therefore construct the conditioning pool from two complementary sources, as illustrated in Figure~\ref{fig:fig1}c. Known patterns are retrieved directly from the training set. For unknown patterns, the number of independent orbits assigned to each Wyckoff position is sampled from its empirical training set distribution within the selected space group, and the sampled counts are assembled into an ordered sequence. This construction retains the observed occupancy support of each position while allowing new combinations across positions; fixed sites inherit a maximum occupancy of one independent orbit. Empty patterns, patterns already present in training, and duplicates of accepted patterns are rejected. The retained patterns condition the same trained model without retraining, extending generation beyond observed joint orbit combinations while maintaining crystallographic occupancy constraints. These patterns provide a data-informed starting point; the validity and energetics of their generated chemical and structural realizations are assessed subsequently. Detailed sampling procedures and network implementation are provided in Supplementary Note 1 and Methods, respectively.

\subsection*{Generation performance based on known and unknown patterns}

To evaluate the quality of ST2M generated crystals, we use the MP20 dataset derived from Materials Project~\cite{bib35}, which yield 43,552 structures after symmetry analysis and reconstruction checks~\cite{bib36,bib37}. Prior to the experiments, to establish a testing basis for evaluating the generated structures, we systematically characterize the chemical, electronic, energetic, and crystallographic distributions of the reference data. (see Supplementary Note 2 and Supplementary Figures 1--3). Additionally, generation is restricted to the 20 most populated space groups, due to the uneven distribution of structures (Supplementary Figure 4 and 5).

Before examining energetic stability, we benchmark the geometric and compositional plausibility of ST2M outputs against established crystal generators, aiming to provide an initial assessment of sample quality (see Supplementary Table 1). As results show, ST2M achieves structural and compositional validities of 96.37\% and 84.04\%, respectively. Its compositional validity exceeds that of DiffCSP, DiffCSP++, and SymmCD, while its structural validity lies within the range of the compared methods. Removing the degree-of-freedom (DOF) masks lowers these values to 92.12\% and 76.74\%, corresponding to decreases of 4.25 and 7.30 percentage points, and increases all four distributional distances. The ablation therefore supports the role of masking in improving both initial sample validity and agreement with the reference data.

\begin{figure}[t]
    \centering
    \includegraphics[width=\textwidth]{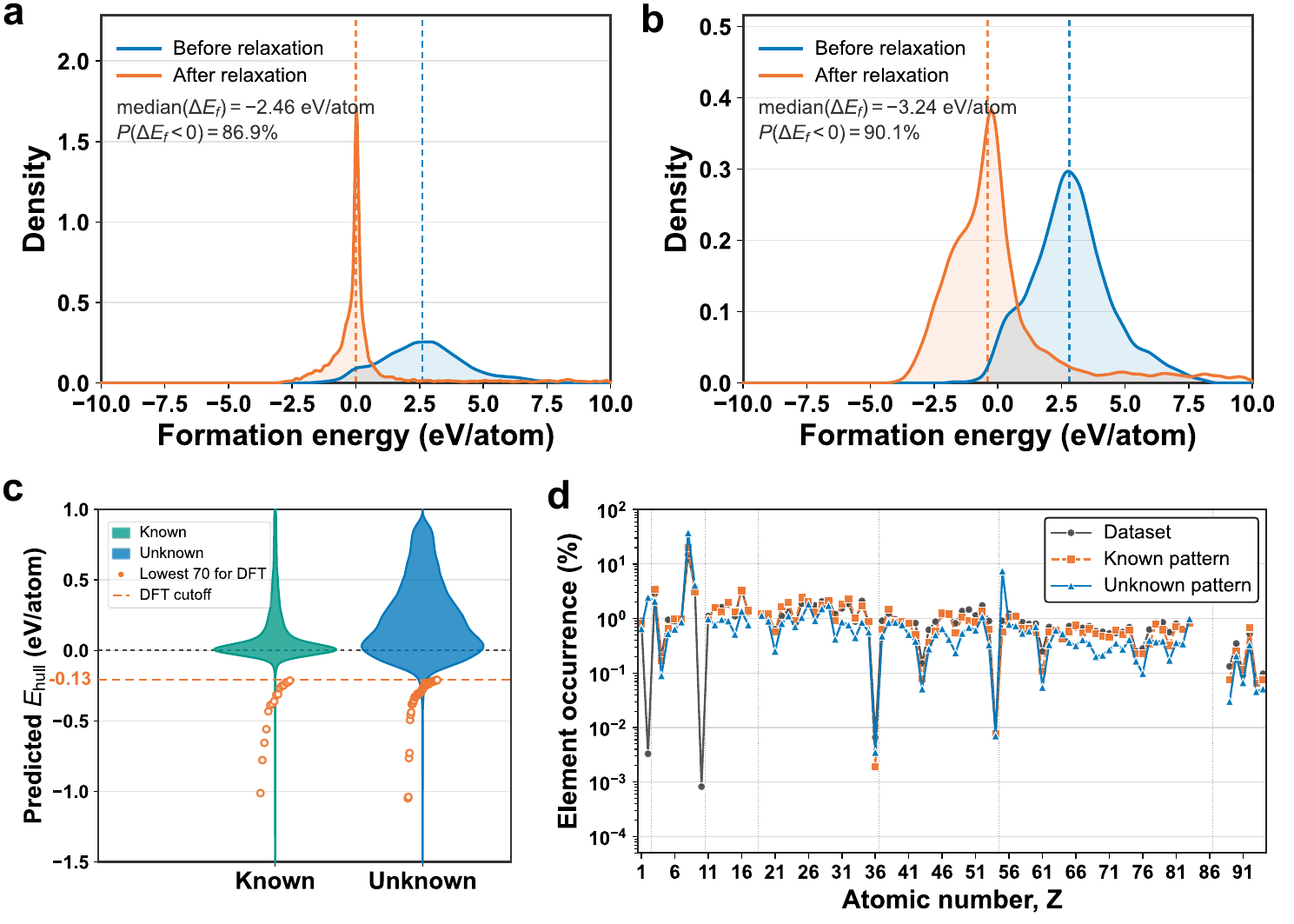}
    \caption{\textbf{Predicted energy distributions and element occurrence frequencies of materials generated by ST2M.} \textbf{a,b}, Predicted formation energy distributions before (blue) and after (orange) relaxation for structures conditioned on known (\textbf{a}) and unknown (\textbf{b}) Wyckoff patterns. Each group contains 9,760 generated structures. Vertical dashed lines mark the distribution medians. With $\Delta E_f=E_f^{\mathrm{relaxed}}-E_f^{\mathrm{initial}}$, the median per-structure energy changes are $-2.46$ and $-3.24$~eV/atom, and 86.9\% and 90.1\% of the evaluated structures exhibit an energy decrease, respectively. \textbf{c}, Distributions of predicted $E_{\mathrm{hull}}$ for known (teal) and unknown (blue) patterns. Here, $E_{\mathrm{hull}}$ denotes the signed energy difference per atom between a candidate and the reference convex hull at the same composition. The reference hull is constructed from Materials Project phases reevaluated using the same surrogate model as the candidates. The candidate is excluded from this reference hull; negative values indicate predicted energies below it. Orange markers highlight the 70 lowest-energy candidates selected from each group for DFT verification. The black dashed line marks $E_{\mathrm{hull}}=0$, and the orange dashed line marks the screening cutoff near $-0.13$~eV/atom. \textbf{d}, Element occurrence frequencies in the training dataset (gray circles) and structures generated from known (orange squares) and unknown (blue triangles) patterns. Each series is normalized by the total number of element occurrences in its corresponding dataset and plotted against atomic number $Z$ on a logarithmic frequency axis.}
    \label{fig:fig2}
\end{figure}

Subsequently, we compare structures generated under known and unknown Wyckoff patterns to determine how extending generation beyond observed orbit combinations affects the energetic quality of ST2M outputs. For each condition, 9,760 structures are generated using the same pretrained model, and their formation energies are evaluated before and after relaxation using surrogate model~\cite{bib16}. Figures~\ref{fig:fig2}a,b show that relaxation shifts both formation energy distributions toward lower energies. The median per-structure energy changes are $-2.46$ and $-3.24$~eV/atom for known and unknown patterns, respectively, while 86.9\% and 90.1\% of the evaluated structures exhibit an energy decrease. These results indicate that the improvement extends across most of each generated population. Comparing the relaxed distributions reveals a further distinction: structures generated from known patterns are sharply concentrated near zero formation energy, whereas those generated from unknown patterns exhibit a broader distribution with a persistent high-energy tail. The larger energy reduction in the unknown pattern group therefore does not imply more favorable final energetics, as it also reflects greater scope for relaxation from the initial configurations. Together, these observations show that structural relaxation improves the predicted energetics under both conditioning regimes, while generation from unknown patterns produces more variable energetic outcomes. Assessing whether the resulting low-energy candidates are stable against competing phases requires the subsequent convex-hull analysis.

In order to evaluate the generation quality of ST2M and determine the candidates for further DFT verification, the relaxed structures are retained and next contrasted against competing phases. We calculate the $E_{\mathrm{hull}}$ by surrogate model and compare the two generated populations in Figure~\ref{fig:fig2}c. Structures generated from known patterns are concentrated near the reference hull, whereas those generated from unknown patterns exhibit a broader distribution extending to higher energies. Approximately, 2,449 and 1,458 relaxed structures are classified as stable out of the 9,760 materials generated under known and unknown patterns, respectively. This comparison indicates that the broader structural exploration enabled by unknown patterns is accompanied by less consistent energetic favorability relative to competing phases. Nevertheless, both populations contain a low-energy tail from which we select the 70 candidates with the lowest predicted $E_{\mathrm{hull}}$ in each group, highlighted by the orange markers. The screening region is indicated by the dashed cutoff near $-0.13$~eV/atom. These selected subsets provide promising candidates from both pattern conditions for testing whether the predicted energetic advantage persists under DFT evaluation.

In addition to the thermodynamic properties, the element distribution of generated materials is also a crucial indicator of wheather the model has learned the characteristics of real materials. We examine the elemental coverage generation under known and unknown Wyckoff patterns, compared with the training data. Element occurrence frequencies are compiled for the training set and both generated populations and compared in Figure~\ref{fig:fig2}d. Both generated groups reproduce many of the dominant peaks and troughs of the reference distribution across the periodic table. Relative to the known pattern group, the unknown pattern group exhibits larger deviations for several elements while retaining broad elemental coverage. This behavior suggests that changing the orbit combinations modifies relative elemental preferences without confining generation to a narrow subset of elements. ST2M therefore retains substantial chemical coverage when generation extends beyond the Wyckoff patterns observed during training.

Ultimately, practical crystal exploration depends on the yield of stable, unique, and novel (S.U.N.) candidates, because a higher yield provides more distinct, previously unexplored candidates for subsequent validation from a given number of generated structures~\cite{bib39}. We therefore compare ST2M with the baseline generators using the relaxed S.U.N. rates in Table~\ref{tab:tab1}. Under known pattern conditioning, ST2M achieves the highest rate of 23.86\%, exceeding the strongest baseline, SymmCD at 14.19\%, by 68.15\% in relative terms. Moreover, 95.1\% of its relaxed structures classified as stable also satisfy the uniqueness and novelty criteria, indicating that the improved stable yield largely translates into new candidates rather than repeated or training-set structures. Under unknown pattern conditioning, the S.U.N. rate remains 14.01\%, close to SymmCD, with 93.8\% of the predicted stable subset retained after the additional checks. Thus, exploration beyond observed orbit combinations remains productive despite its lower stable fraction. ST2M delivers the highest S.U.N. yield in this benchmark under known pattern conditioning, increasing the supply of distinct, potentially stable materials per generated sample and supporting its practical use in structure exploration.

\begin{table}[h]
    \centering
    \caption{Stability and S.U.N. statistics of generated structures. All percentages are relative to the full generated population.}
    \label{tab:tab1}
    \begin{tabular}{lccc}
        \toprule
        Method & Initial stable & Relaxed stable & Relaxed S.U.N. \\
        \midrule
        CDVAE        & 0.32\%  & 6.81\%  & 5.09\%  \\
        DiffCSP      & 8.26\%  & 13.94\% & 12.26\% \\
        DiffCSP++    & 7.79\%  & 13.81\% & 12.85\% \\
        SymmCD       & 9.21\%  & 15.48\% & 14.19\% \\
        ST2M$_{\mathrm{known}}$       & 17.02\% & 25.09\% & 23.86\% \\
        ST2M$_{\mathrm{unknown}}$ & 5.59\% & 14.94\% & 14.01\% \\
        \bottomrule
    \end{tabular}
\end{table}

\subsection*{Computational validation and characterization of generated crystals}

\begin{figure}[t]
    \centering
    \includegraphics[width=\textwidth]{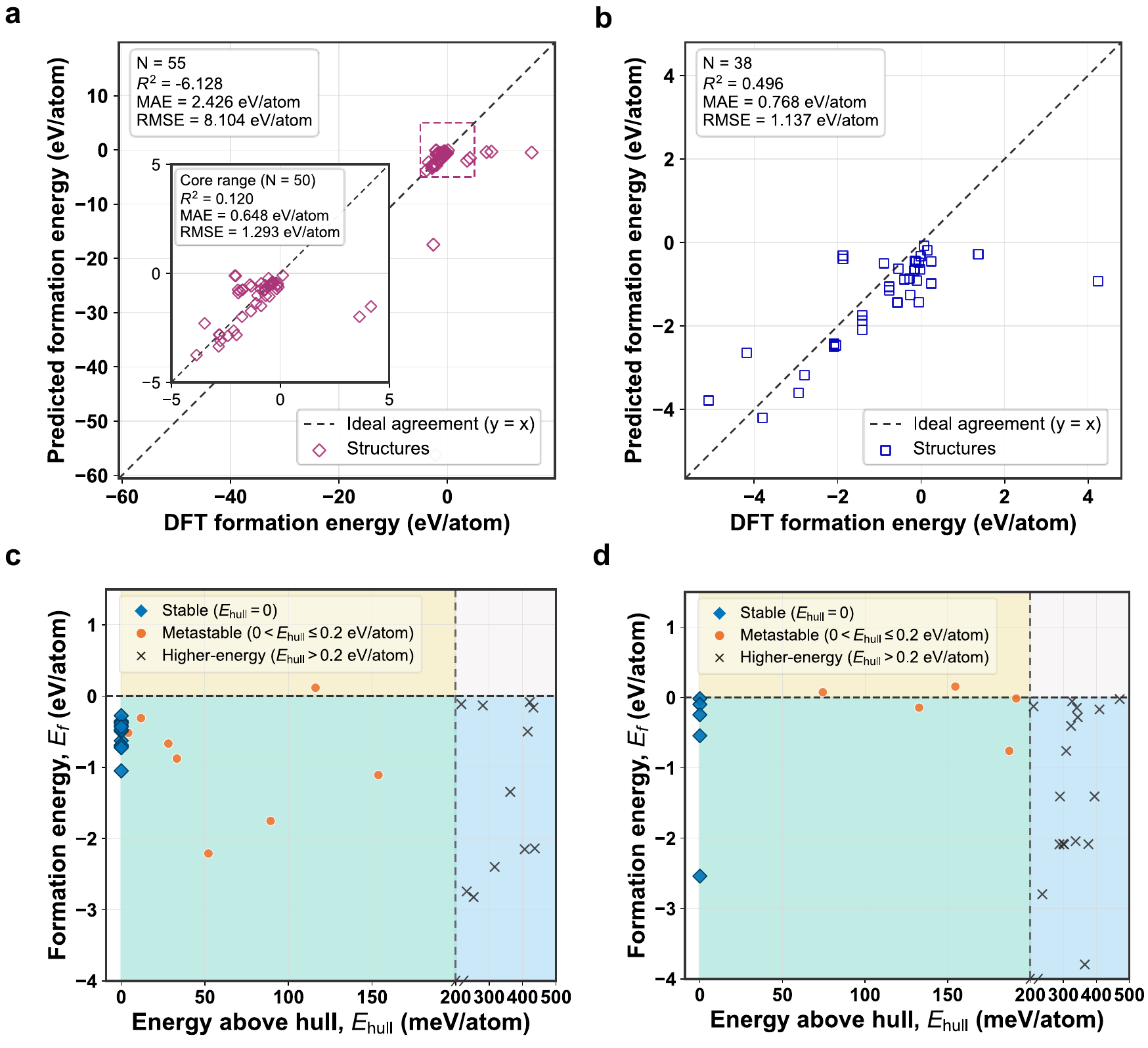}
    \caption{\textbf{DFT validation of ST2M-generated candidates and comparison with surrogate predictions.} \textbf{a,b}, Surrogate predicted versus DFT formation energies for 55 candidates generated from known patterns (\textbf{a}) and 38 from unknown patterns (\textbf{b}). Dashed lines indicate ideal agreement, $y=x$. The inset in \textbf{a} enlarges the central energy range and reports statistics for the 50 structures within that range. $R^2$, MAE, and RMSE denote the coefficient of determination, mean absolute error, and root-mean-square error, respectively; energy errors are in eV/atom. \textbf{c,d}, DFT formation energies, $E_f$, versus energies above the convex hull, $E_{\mathrm{hull}}$, for candidates from known (\textbf{c}) and unknown (\textbf{d}) patterns. Blue diamonds indicate stable candidates with $E_{\mathrm{hull}}=0$, orange circles indicate the adopted metastable range, $0<E_{\mathrm{hull}}\leq0.2$~eV/atom, and gray crosses indicate higher-energy candidates. Horizontal axes use meV/atom and are compressed above 200~meV/atom. Dashed lines at $E_f=0$ and $E_{\mathrm{hull}}=200$~meV/atom delimit the shaded regions. The DFT assessment identifies 16 and five thermodynamically stable structures from the known- and unknown pattern shortlists, respectively.}
    \label{fig:fig3}
\end{figure}

Apart from the solely ML predicted results, the further DFT verification is also necessary, because a favorable surrogate energy does not by itself establish the physical plausibility of a generated structure. We therefore compare surrogate predictions with DFT formation energies to assess whether ST2M generated candidates are amenable to large-scale surrogate assisted screening. Of the 70 candidates selected from each pattern group, complete paired results are available for 55 known pattern and 38 unknown pattern structures. Figures~\ref{fig:fig3}a,b show closer energy agreement for the central subset of 50 known pattern structures, with a mean absolute error of 0.648~eV/atom, whereas several large discrepancies increase the error for the full known pattern set to 2.426~eV/atom. The unknown pattern candidates exhibit a mean absolute error of 0.768~eV/atom and $R^2=0.496$. In addition, DFT yields negative formation energies for 49 of the 55 known pattern candidates and 32 of the 38 unknown pattern candidates, corresponding to 89.1\% and 84.2\% of the evaluated structures, respectively (Figures~\ref{fig:fig3}a,b). These results show that most evaluated ST2M candidates are energetically favored relative to their elemental reference states, providing a formation energy assessment of the generated structures in addition to the surrogate--DFT comparison. The agreement within these subsets supports preliminary surrogate based prioritization of the high throughout exploration of the generated configurations. The agreements shown in Figures~\ref{fig:fig3}a,b also demonstrate that materials generated by ST2M are close to the real crystal structures that used for the training of the ML potential, as both surrogate model and DFT calculation yield the similar results. 

After assessing the validation and stability based on formation energy, whether the screened structures resist decomposition into competing phases is also the decisive test of their thermodynamic viability. We evaluate candidate structures and Materials Project reference phases using a consistent DFT protocol~\cite{bib40,bib41,bib42} and construct the corresponding convex hulls (see Supplementary Note 3). Figures~\ref{fig:fig3}c,d show that negative formation energies can coexist with appreciable energies above the hull, so formation relative to the elements alone does not establish stability against competing phases. DFT identifies 16 stable structures from known patterns and five from unknown patterns, corresponding to 22.9\% and 7.1\% of the original 70-structure shortlists, respectively. An additional eight known pattern and five unknown pattern candidates lie in the adopted metastable region of Figures~\ref{fig:fig3}c,d. Together with the stable structures, these give 24 of 55 (43.6\%) and 10 of 38 (26.3\%) evaluated candidates in the stable or metastable categories, respectively, showing that both conditions recover additional low-energy competitors while known patterns yield a larger fraction of candidates on or near the DFT hull. The higher yield for known patterns follows the screening trends in Table~\ref{tab:tab1}. The five stable unknown pattern candidates nevertheless demonstrate that ST2M can explore previously unobserved orbit combinations while producing structures that remain energetically favorable against the reference phases included in the DFT assessment.

Thermodynamic stability against competing phases does not establish local stability against lattice distortions. We therefore examine the 21 retained candidates using phonon dispersions predicted by the benchmarked PhononBench workflow from MatterSim forces and finite-displacement calculations in Phonopy~\cite{bib43,bib44,bib45}. All exhibit no appreciable imaginary branches along the sampled paths, supporting harmonic dynamical stability within the employed machine learning potential (Supplementary Figures 9--24 for known patterns and 25--29 for unknown patterns). Notably, the elemental As candidate adopts space group $P2_1/c$ (No.~14) and lies approximately 18~meV/atom below the lowest-energy As reference retrieved from Materials Project and recalculated using the same DFT protocol. Its favorable energy and predicted phonon stability identify a candidate As allotrope that merits further investigation (see Supplementary Figure 26).

To acquire a comprehensive analysis, the retained candidates that pass the S.U.N. test are compared to determine whether prescribed crystallographic conditions permit diversity in composition, atomic arrangement, and electronic behavior. Table~\ref{tab:tab2} shows that the 21 structures span 12 space groups and elemental, binary, ternary, and quaternary compositions, with formation energies ranging from $-3.892$ to $-0.018$~eV/atom and signed reference-hull energy differences ranging from $-3.257$ to $-0.012$~eV/atom. All retain their space group numbers and occupied Wyckoff patterns after relaxation, indicating compatibility between the prescribed assignments and the relaxed structures. The two $\mathrm{ThMn}_{2}\mathrm{SiC}$ configurations share space group 63 and pattern $[a,c,c,f]$, yet configuration II is lower in energy by 16~meV/atom, demonstrating distinct relaxed arrangements within a common composition and orbit sequence (Supplementary Figures 10 and 17). The five unknown pattern candidates, $\mathrm{C}_{3}\mathrm{La}_{2}\mathrm{Mo}_{2}$, $\mathrm{As}$, $\mathrm{Ag}_{3}\mathrm{Au}_{2}\mathrm{Pd}_{3}$, $\mathrm{Ca}_{2}\mathrm{Rh}$, and $\mathrm{CrF}_{3}$, extend this set across different compositions and orbit combinations. Electronic calculations identify two semiconductors: $\mathrm{Cl}_{6}\mathrm{Cs}_{2}\mathrm{LiRh}$ has an indirect PBE gap of 1.88~eV, while $\mathrm{CrF}_{3}$ has a direct gap of 1.13~eV with both band edges at $\Gamma$ along the sampled path (Supplementary Figure 29c). The remaining 19 structures are metallic. The retained set therefore encompasses distinct electronic behaviors under both pattern conditions without explicit electronic property conditioning.

\begin{table}[htbp]
    \centering
    \caption{Chemical compositions, crystallographic assignments, lattice parameters, band gaps, and DFT energetics of the 21 retained candidates.}
    \label{tab:tab2}
    \setlength{\tabcolsep}{1.5pt}
    \scriptsize
    \begin{tabular}{lccccccc}
        \toprule
        Formula & Pattern type & Space group & Wyckoff pattern & Lattice parameters\footnotemark[1] & Band gap\footnotemark[2] & $E_f$ & $\Delta E_{\mathrm{hull}}^{\mathrm{ref}}$\footnotemark[3] \\
        & & & & (\AA) & (eV) & (eV/atom) & (eV/atom) \\
        \midrule
        $\mathrm{Cl}_{6}\mathrm{Cs}_{2}\mathrm{Li}\mathrm{Rh}$ & known & 164 ($P\bar{3}m1$) & $[a,b,d,i]$ & (7.00, 7.00, 6.34) & 1.88 & -1.475 & -0.037 \\
        $\mathrm{Th}\mathrm{Mn}_{2}\mathrm{Si}\mathrm{C}$\textsuperscript{I} & known & 63 ($Cmcm$) & $[a,c,c,f]$ & (3.89, 10.66, 6.96) & Metal & -0.461 & -0.076 \\
        $\mathrm{Lu}\mathrm{Ag}$ & known & 221 ($Pm\bar{3}m$) & $[a,b]$ & (3.56, 3.56, 3.56) & Metal & -0.359 & -0.103 \\
        $\mathrm{Sr}\mathrm{Sn}_{2}\mathrm{Pd}_{2}$ & known & 129 ($P4/nmm$) & $[a,b,c,c,c]$ & (4.71, 4.71, 11.39) & Metal & -0.712 & -0.117 \\
        $\mathrm{Ce}\mathrm{Fe}_{2}\mathrm{Si}\mathrm{C}$ & known & 63 ($Cmcm$) & $[a,c,c,f]$ & (3.79, 11.30, 6.91) & Metal & -0.671 & -0.292 \\
        $\mathrm{Pm}\mathrm{Tl}$ & known & 221 ($Pm\bar{3}m$) & $[a,b]$ & (7.84, 7.84, 7.84) & Metal & -0.384 & -2.528 \\
        $\mathrm{Pm}\mathrm{Pt}_{2}$ & known & 63 ($Cmcm$) & $[a,c,c]$ & (4.17, 10.00, 5.54) & Metal & -2.556 & -2.556 \\
        $\mathrm{Ga}\mathrm{Pd}\mathrm{Th}$ & known & 189 ($P\bar{6}2m$) & $[b,c,f,g]$ & (6.98, 6.98, 4.14) & Metal & -0.820 & -0.055 \\
        $\mathrm{Th}\mathrm{Mn}_{2}\mathrm{Si}\mathrm{C}$\textsuperscript{II} & known & 63 ($Cmcm$) & $[a,c,c,f]$ & (3.67, 10.62, 6.84) & Metal & -0.477 & -0.092 \\
        $\mathrm{Cs}\mathrm{P}_{2}\mathrm{Os}_{2}$ & known & 139 ($I4/mmm$) & $[a,d,e]$ & (4.03, 4.03, 15.58) & Metal & -0.369 & -0.057 \\
        $\mathrm{Ag}\mathrm{Nd}\mathrm{Sb}_{2}$ & known & 129 ($P4/nmm$) & $[a,b,c,c]$ & (4.51, 4.51, 10.35) & Metal & -1.489 & -0.795 \\
        $\mathrm{Pd}\mathrm{Si}\mathrm{Ta}$ & known & 62 ($Pnma$) & $[c,c,c]$ & (6.22, 3.84, 7.12) & Metal & -0.292 & -0.012 \\
        $\mathrm{Al}\mathrm{C}\mathrm{Hf}_{2}$ & known & 194 ($P6_{3}/mmc$) & $[a,d,f]$ & (3.39, 3.39, 13.08) & Metal & -0.701 & -0.027 \\
        $\mathrm{Ba}\mathrm{Tl}$ & known & 221 ($Pm\bar{3}m$) & $[a,b]$ & (4.32, 4.32, 4.32) & Metal & -0.403 & -0.097 \\
        $\mathrm{Gd}\mathrm{Pd}\mathrm{Pb}$ & known & 189 ($P\bar{6}2m$) & $[b,c,f,g]$ & (7.53, 7.53, 3.85) & Metal & -3.892 & -3.257 \\
        $\mathrm{Nb}\mathrm{Si}\mathrm{Mo}$ & known & 189 ($P\bar{6}2m$) & $[b,c,f,g]$ & (6.84, 6.84, 3.59) & Metal & -0.544 & -0.038 \\
        $\mathrm{C}_{3}\mathrm{La}_{2}\mathrm{Mo}_{2}$ & unknown & 12 ($C2/m$) & $[c,i,i,i]$ & (11.74, 3.81, 5.87) & Metal & -0.161 & -0.034 \\        
        $\mathrm{As}$ & unknown & 14 ($P2_1/c$) & $[e]$ & (4.67, 5.86, 5.67) & Metal & -0.018 & -0.018 \\
        $\mathrm{Ag}_{3}\mathrm{Au}_{2}\mathrm{Pd}_{3}$ & unknown & 166 ($R\bar{3}m$) & $[a,b,d,e]$ & (5.93, 5.93, 13.40) & Metal & -0.100 & -0.036 \\
        $\mathrm{Ca}_{2}\mathrm{Rh}$ & unknown & 221 ($Pm\bar{3}m$) & $[b,c,g]$ & (7.31, 7.31, 7.31) & Metal & -0.246 & -0.014 \\
        $\mathrm{Cr}\mathrm{F}_{3}$ & unknown & 15 ($C2/c$) & $[d,e,f]$ & (8.32, 6.15, 6.65) & 1.13 & -2.540 & -0.021 \\
        \bottomrule
    \end{tabular}
\footnotetext[1]{Lattice parameters are listed in the order $a$, $b$, $c$.}
\footnotetext[2]{Metal denotes a metallic electronic structure; band gaps are calculated using PBE.}
\footnotetext[3]{$\Delta E_{\mathrm{hull}}^{\mathrm{ref}}$ is the signed energy difference relative to the reference convex hull, excluding the candidates; negative values indicate energies below this reference hull.}
\end{table}

\begin{figure}[t]
    \centering
    \includegraphics[width=\textwidth]{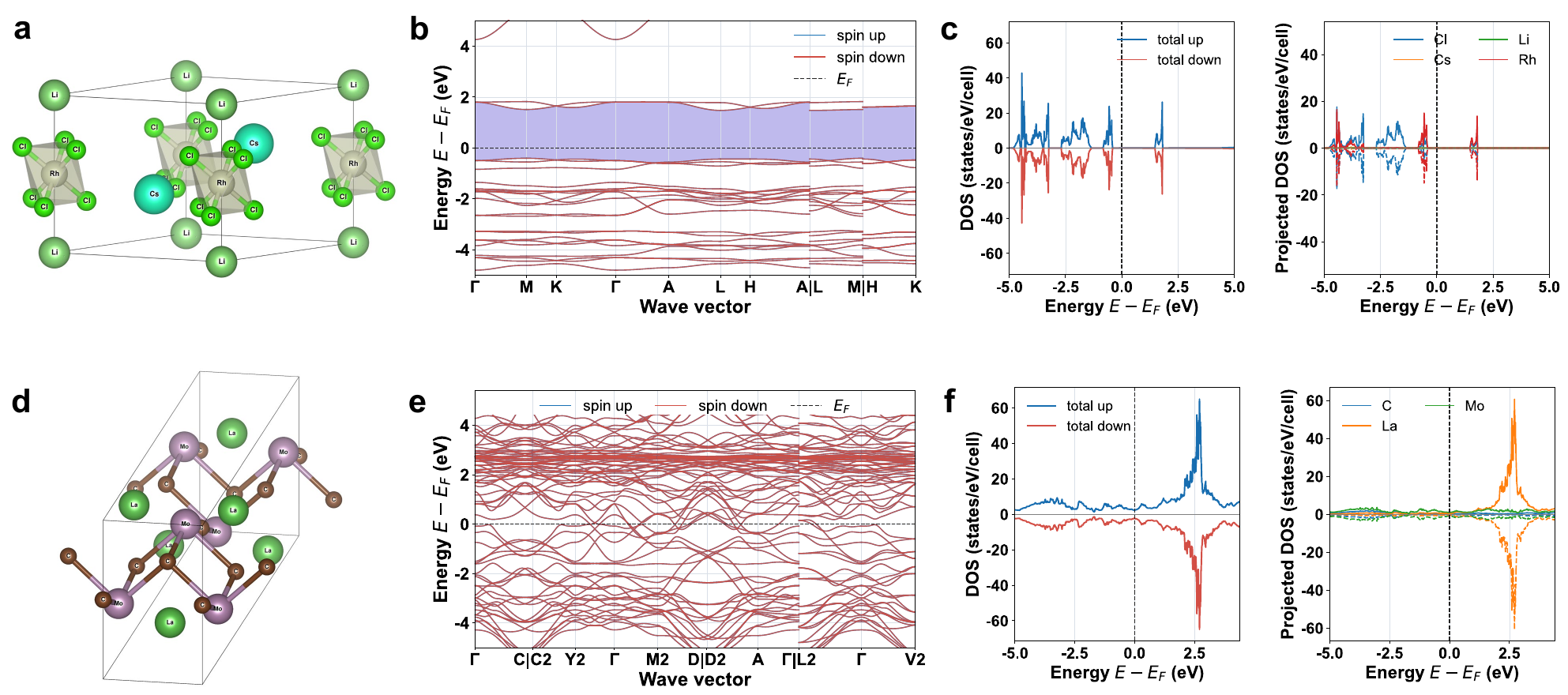}
    \caption{\textbf{Crystal structures and DFT electronic properties of representative ST2M-generated candidates.} \textbf{a--c}, $\mathrm{Cl}_{6}\mathrm{Cs}_{2}\mathrm{LiRh}$ generated from a known Wyckoff pattern: crystal structure (\textbf{a}), spin-resolved electronic band structure (\textbf{b}), and total DOS (left) and element projected DOS (right; \textbf{c}). \textbf{d--f}, Corresponding results for $\mathrm{C}_{3}\mathrm{La}_{2}\mathrm{Mo}_{2}$ generated from an unknown Wyckoff pattern: crystal structure (\textbf{d}), electronic band structure (\textbf{e}), and total and element projected DOS (\textbf{f}). Energies are referenced to the Fermi level, $E_F$, indicated by dashed lines. In the total DOS plots, spin-up and spin-down contributions are plotted above and below zero, respectively; the projected DOS uses the same sign convention. $\mathrm{Cl}_{6}\mathrm{Cs}_{2}\mathrm{LiRh}$ exhibits a finite band gap, whereas $\mathrm{C}_{3}\mathrm{La}_{2}\mathrm{Mo}_{2}$ shows bands crossing $E_F$ and a finite DOS at $E_F$.}
    \label{fig:fig4}
\end{figure}

Specifically, atomic coordination and electronic states are examined for two representative candidates to characterize the materials obtained under known and unknown pattern conditions. Figure~\ref{fig:fig4} contrasts the semiconducting $\mathrm{Cl}_{6}\mathrm{Cs}_{2}\mathrm{LiRh}$ structure with metallic $\mathrm{C}_{3}\mathrm{La}_{2}\mathrm{Mo}_{2}$. The DFT-relaxed known pattern candidate in Figure~\ref{fig:fig4}a adopts trigonal $P\bar{3}m1$ symmetry, with $a=b=7.355$~\AA{} and $c=6.180$~\AA{}. Li and Rh occupy the fixed $a$ and $b$ sites, whereas Cs and Cl occupy the $d$ and $i$ orbits. Six symmetry-equivalent Cl neighbors surround each Rh at 2.371~\AA{}, forming slightly distorted $\mathrm{RhCl}_{6}$ octahedra that share no Cl vertices. The indirect PBE gap of 1.88~eV has its valence-band maximum at $M$ and conduction-band minimum at $L$ (Figure~\ref{fig:fig4}b). The DOS vanishes at the Fermi level, with substantial Cl contributions to the occupied states below the gap and Rh contributions to the low-lying unoccupied states (Figure~\ref{fig:fig4}c). The unknown pattern candidate in Figure~\ref{fig:fig4}d instead adopts monoclinic $C2/m$ symmetry, with $a=11.644$~\AA{}, $b=4.127$~\AA{}, $c=5.596$~\AA{}, and $\beta=114.788^{\circ}$. Its 14-atom conventional cell contains two formula units, with two inequivalent C sites on the $c$ and $i$ orbits and La and Mo on separate $i$ orbits. Each Mo has four nearby C neighbors at 2.055--2.099~\AA{}. Bands cross the Fermi level (Figure~\ref{fig:fig4}e), where the spin-summed DOS reaches 4.50~states\,eV$^{-1}$\,cell$^{-1}$ and Mo provides the largest elemental contribution; La dominates the unoccupied DOS peak approximately 2.71~eV above $E_F$ (Figure~\ref{fig:fig4}f). These cases connect the generated orbit arrangements to distinct local coordination and electronic characteristics within the validation workflow (see Supplementary Figure 6 and 8).

\section*{Discussion}

ST2M combines explicit crystallographic constraints with flexible generation of elemental identities and structural parameters, improving the yield of physically plausible and diverse crystal candidates. Its DOF masks restrict continuous diffusion to independent coordinate and lattice variables, removing symmetry-fixed quantities from the denoising targets. This reduces redundant learning demands and directs model capacity toward variables that determine distinct structures. Wyckoff pattern conditioning complements this mechanism by defining compatible occupied orbit arrangements while leaving elemental assignments and symmetry allowed geometry open to exploration. The resulting search space therefore retains chemical and structural flexibility within a prescribed symmetry framework. The improvements in validity and distributional agreement over the unmasked model support this design, while the relaxed S.U.N. rate of 23.86\% under known pattern conditioning demonstrates a higher yield of useful candidates per generated sample than the compared baselines. The 21 DFT validated stable candidates spanning 12 space groups, including five generated from unknown patterns, further show that these constraints accommodate diverse, energetically viable structures beyond observed orbit combinations.

The current framework nevertheless generates crystal structures without conditioning on target properties or prescribed elemental combinations. It therefore does not yet support inverse design in which a desired material property guides structure generation. Incorporating property conditioning into the denoising process would extend ST2M toward targeted discovery, allowing the search to prioritize structures with specified functional characteristics. A second priority is element-set conditioning, which would restrict elemental assignments to a chosen chemical system and enable focused exploration of experimentally or technologically relevant element combinations. These extensions should preserve the existing symmetry and orbit constraints while providing direct control over the chemical and functional scope of generation. The lower DFT-stable yield for unknown patterns also motivates stronger stability aware guidance, so that extending structural diversity more consistently translates into energetically viable candidates. Together, these developments would advance ST2M from symmetry controlled structure generation toward targeted inverse materials design.

\section*{Methods}

\subsection*{Symmetry extraction and compact Wyckoff representation}

Each input structure is converted to a standardized conventional cell before symmetry analysis. The occupied crystallographic orbits are then extracted according to the identified space group $G$. For orbit $i$, we retain its element label $z_{0,i}$, Wyckoff letter $w_i$, multiplicity, number of positional DOF, and one representative fractional coordinate $\mathbf{u}_{0,i}$. Applying the space group operations associated with $(G,w_i)$ to this representative recovers all symmetry-equivalent atoms in the orbit. Only structures whose encoded representation reconstructs the original standardized structure are retained.

For a crystal containing $N$ independent occupied orbits, the categorical labels and representative coordinates satisfy $\mathbf{Z}_0\in\{1,\ldots,100\}^{N}$ and $\mathbf{U}_0\in\mathbb{R}^{N\times3}$. Category 0 is reserved as the absorbing and vacancy class. The orbit sequence is sorted first by Wyckoff letter and then, for repeated letters, lexicographically by the representative coordinates mapped to $[0,1)$. This deterministic ordering reduces representational ambiguity while allowing multiple independent orbits to occupy the same non-fixed Wyckoff type.

Consistent with the notation in Results, $\mathbf{Z}_t$ denotes the categorical element state, while $\mathbf{U}_t$ and $\boldsymbol{\ell}_t$ denote the continuous coordinate and lattice states, respectively. The ordered Wyckoff sequence is $\mathcal{W}=(w_1,\ldots,w_N)=\mathcal{W}(\mathbf{n})$, where $\mathbf{n}=(n_w)_{w\in\mathcal{A}_G}$ counts independent occupied orbits of each Wyckoff type and $N=\sum_{w\in\mathcal{A}_G}n_w$, as defined in Supplementary Note 2.

For each $(G,w_i)$ pair, a binary coordinate mask is defined as
\begin{equation}
    \mathbf{M}^{u}_i=\mathbf{M}^{u}(G,w_i)\in\{0,1\}^{3},
    \qquad
    \|\mathbf{M}^{u}_i\|_1=d(G,w_i),
    \label{eq:method_coordinate_mask}
\end{equation}
where $d(G,w_i)\in\{0,1,2,3\}$ is the number of independent positional parameters. For example, a Wyckoff coordinate of the form $(x,1/2,1/4)$ has $\mathbf{M}^{u}_i=(1,0,0)$. Only $x$ is stored as a generative DOF; the constrained components are recovered from the Wyckoff expression. Fixed positions have $\mathbf{M}^{u}_i=\mathbf{0}$.

The lattice is represented by
\begin{equation}
    \boldsymbol{\ell}_0=
    \left(\log a,\log b,\log c,
    \cos\alpha,\cos\beta,\cos\gamma\right).
    \label{eq:method_lattice_representation}
\end{equation}
The logarithmic lengths remain unconstrained during diffusion but yield positive edge lengths after exponentiation. A space group-dependent mask $\mathbf{M}^{\ell}_G\in\{0,1\}^{6}$ retains only the independent lattice variables of the associated crystal system. For example, tetragonal structures generate $\log a$ and $\log c$, whereas $b=a$ and all angles are fixed; cubic structures generate only one length. For trigonal and hexagonal settings, $a=b$, $\alpha=\beta=90^{\circ}$, and $\gamma=120^{\circ}$ are imposed during reconstruction.

\subsection*{Coupled diffusion and dual-branch denoising in ST2M}

\noindent\textit{Forward diffusion.} For every crystal, a diffusion step $t\sim\mathcal{U}\{0,\ldots,T-1\}$ is sampled and broadcast to all its orbit tokens. ST2M applies coordinated but modality-specific forward processes to the categorical element identities and continuous structural variables. Element types follow an absorbing-state categorical process with transition matrix~\cite{bib46}
\begin{equation}
    \mathbf{Q}_t=(1-\beta^{D}_t)\mathbf{I}
    +\beta^{D}_t\mathbf{1}\mathbf{e}_0^{\mathsf T},
    \qquad
    \overline{\mathbf{Q}}_t=\prod_{s=0}^{t}\mathbf{Q}_s,
    \label{eq:method_d3pm_transition}
\end{equation}
where $\mathbf{e}_0$ is concentrated on class 0. If $\mathbf{d}_{0,i}$ is the one-hot encoding of $z_{0,i}$, the forward marginal is
\begin{equation}
    q(z_{t,i}\mid z_{0,i})=
    \operatorname{Cat}\!\left(
    \mathbf{d}_{0,i}^{\mathsf T}\overline{\mathbf{Q}}_t
    \right).
    \label{eq:method_d3pm_forward}
\end{equation}
The terminal transition maps the element state to the absorbing class. The discrete branch predicts the clean distribution $p_{\theta_D}(z_{0,i}\mid\mathbf{Z}_t,\mathbf{U}_t,\boldsymbol{\ell}_t,t,G,\mathcal{W})$, which is combined with the analytic D3PM posterior to obtain the reverse transition.

The representative coordinates and lattice parameters follow masked Gaussian processes~\cite{bib47},
\begin{equation}
\begin{aligned}
    \mathbf{u}_{t,i}
    &=\mathbf{M}^{u}_i\odot\left(
    \sqrt{\bar{\alpha}^{C}_t}\,\mathbf{u}_{0,i}
    +\sqrt{1-\bar{\alpha}^{C}_t}\,\boldsymbol{\epsilon}^{u}_i
    \right),
    &\boldsymbol{\epsilon}^{u}_i&\sim\mathcal{N}(\mathbf{0},\mathbf{I}),
    \\
    \boldsymbol{\ell}_{t}
    &=\mathbf{M}^{\ell}_G\odot\left(
    \sqrt{\bar{\alpha}^{C}_t}\,\boldsymbol{\ell}_{0}
    +\sqrt{1-\bar{\alpha}^{C}_t}\,\boldsymbol{\epsilon}^{\ell}
    \right),
    &\boldsymbol{\epsilon}^{\ell}&\sim\mathcal{N}(\mathbf{0},\mathbf{I}),
\end{aligned}
\label{eq:method_continuous_forward}
\end{equation}
where $\bar{\alpha}^{C}_t$ is the cumulative continuous noise schedule and $\odot$ denotes element-wise multiplication. Masked components remain zero at every diffusion step and do not become learning targets.

\noindent\textit{Dual-branch Wyckoff-conditioned denoiser.} The reverse dynamics are parameterized by two independently Transformer encoders based on the self-attention architecture~\cite{bib48}. The discrete branch recovers element identities, whereas the continuous branch predicts the noise applied to the Wyckoff free coordinates and lattice variables. Both branches observe the complete noisy state and the same crystallographic conditions, enabling bidirectional chemical--geometric conditioning without sharing network parameters. Here, the subscripts $D$ and $C$ label the discrete and continuous branches, respectively. For branch $X\in\{D,C\}$, the input embedding of orbit $i$ is
\begin{equation}
\begin{split}
    \mathbf{h}^{X,(0)}_i={}&
    \phi^{X}_{D}(\mathbf{d}_{t,i})
    +\phi^{X}_{u}(\mathbf{u}_{t,i})
    +\phi^{X}_{\ell}(\boldsymbol{\ell}_{t,b(i)})
    +\phi^{X}_{t}\!\left[\operatorname{PE}(t_{b(i)})\right]\\
    &+\mathbf{E}^{X}_{W}(w_i)
    +\mathbf{E}^{X}_{G}(G_{b(i)}),
    \label{eq:method_token_embedding}
\end{split}
\end{equation}
where $b(i)$ indexes the crystal containing token $i$, $N_b$ is the number of independent occupied orbits in crystal $b$, $\operatorname{PE}(\cdot)$ is a sinusoidal time embedding, the $\phi(\cdot)$ functions are learnable multilayer perceptrons, and $\mathbf{E}_{W}$ and $\mathbf{E}_{G}$ are embeddings of the Wyckoff letter and space group. The two branches use the same input variables but do not share their embedding layers, Transformer layers, or output heads.

Packed orbit tokens are processed with a block-diagonal attention mask,
\begin{equation}
    A_{ij}=\begin{cases}
    0, & b(i)=b(j),\\
    -\infty, & b(i)\neq b(j),
    \end{cases}
    \label{eq:method_attention_mask}
\end{equation}
so that all independent orbits within a crystal can interact while tokens from different crystals remain isolated. The discrete branch produces 101 clean-element logits per orbit. The continuous branch produces a three-dimensional coordinate-noise estimate for every orbit. Its hidden states are mean-pooled within each crystal and passed through a lattice head to predict the six-dimensional lattice noise,
\begin{equation}
    \overline{\mathbf{h}}^{C}_b=
    \frac{1}{N_b}\sum_{i:b(i)=b}\mathbf{h}^{C}_i,
    \qquad
    \widehat{\boldsymbol{\epsilon}}^{\ell}_{\theta_C,b}=
    \phi_{\ell}(\overline{\mathbf{h}}^{C}_b).
    \label{eq:method_lattice_head}
\end{equation}

\subsection*{Optimization objectives and crystal reconstruction}

\noindent\textit{Symmetry-masked denoising objectives.} ST2M optimizes the discrete branch with a D3PM objective~\cite{bib46} and the continuous branch with DDPM noise-prediction objectives~\cite{bib47} (Supplementary Figure 7). At $t=0$, the discrete objective is the clean-element cross-entropy. At $t>0$, it is the Kullback--Leibler divergence between the exact categorical posterior and the model posterior,
\begin{equation}
    \mathcal{L}_{\mathrm{disc}}=
    \begin{cases}
    -\log p_{\theta_D}(z_{0,i}\mid\mathbf{Z}_t,\mathbf{U}_t,\boldsymbol{\ell}_t,t,G,\mathcal{W}), & t=0,\\[2pt]
    D_{\mathrm{KL}}\!\left[
    q(z_{t-1,i}\mid z_{t,i},z_{0,i})\,\|\,
    p_{\theta_D}(z_{t-1,i}\mid\mathbf{Z}_t,\mathbf{U}_t,\boldsymbol{\ell}_t,t,G,\mathcal{W})
    \right], & t>0.
    \end{cases}
    \label{eq:method_discrete_loss}
\end{equation}
The loss is averaged over valid occupied orbit tokens. The continuous branch is optimized through masked noise-prediction losses,
\begin{align}
    \mathcal{L}_{\mathrm{coord}}
    &=\frac{
    \sum_i\left\|\mathbf{M}^{u}_i\odot
    (\boldsymbol{\epsilon}^{u}_i-
    \widehat{\boldsymbol{\epsilon}}^{u}_{\theta_C,i})
    \right\|_2^2}
    {\sum_i\|\mathbf{M}^{u}_i\|_1},
    \label{eq:method_coordinate_loss}\\
    \mathcal{L}_{\mathrm{lat}}
    &=\frac{
    \sum_b\left\|\mathbf{M}^{\ell}_{G_b}\odot
    (\boldsymbol{\epsilon}^{\ell}_b-
    \widehat{\boldsymbol{\epsilon}}^{\ell}_{\theta_C,b})
    \right\|_2^2}
    {\sum_b\|\mathbf{M}^{\ell}_{G_b}\|_1}.
    \label{eq:method_lattice_loss}
\end{align}
The denominators count the independent scalar variables contributing to each objective. Consequently, fixed coordinates and symmetry-dependent lattice components neither receive noise nor contribute gradients.

\noindent\textit{Joint optimization and crystal reconstruction.} The DDPM coordinate and lattice losses and the D3PM categorical loss define the continuous and discrete branch objectives,
\begin{equation}
\begin{split}
    \mathcal{L}_{C}&=\lambda_{u}\mathcal{L}_{\mathrm{coord}}
    +\lambda_{\ell}\mathcal{L}_{\mathrm{lat}},\\
    \mathcal{L}_{D}&=\lambda_{D}\mathcal{L}_{\mathrm{disc}},\\
    \mathcal{L}_{\mathrm{total}}&=\mathcal{L}_{C}+\mathcal{L}_{D}.
    \label{eq:method_total_objective}
\end{split}
\end{equation}
Here, $\lambda_u$, $\lambda_{\ell}$, and $\lambda_D$ weight the coordinate, lattice, and categorical denoising losses, respectively. The separate parameter sets are optimized jointly, while each branch receives gradients only from its associated denoising objectives.

At generation time, a space group $G$ and an occupied Wyckoff-orbit sequence $\mathcal{W}$ are drawn from the prescribed Wyckoff pattern pool. ST2M then performs the coupled reverse process from the following categorical and continuous initial states:
\begin{equation}
    \mathbf{Z}_{T}=\mathbf{0},
    \qquad
    \mathbf{U}_{T}\sim\mathcal{N}(\mathbf{0},\mathbf{I})\odot\mathbf{M}^{u},
    \qquad
    \boldsymbol{\ell}_{T}\sim\mathcal{N}(\mathbf{0},\mathbf{I})\odot\mathbf{M}^{\ell}_{G}.
    \label{eq:method_sampling_initialization}
\end{equation}
From $t=T-1$ to 0, the discrete state is sampled from the D3PM model posterior, while the representative coordinates and lattice follow their Gaussian reverse transitions. Class-0 orbit tokens are removed after the final step. The independent lattice variables are transformed back to edge lengths and angles and then completed using the crystal system constraints. Finally, the generated free coordinates are inserted into the corresponding Wyckoff expressions, and the space group operations expand all independent representatives into the full conventional unit cell.

\section*{Acknowledgements}
This work was supported by the National Natural Science Foundation of China (12374057) and the Natural Science Foundation of Sichuan Province (2026NSFSC0030). The work (S.T.) at Los Alamos National Laboratory (LANL) was performed at the Center for Integrated Nanotechnologies (CINT), a U.S. Department of Energy, Office of Science user facility at LANL.

\section*{Contributions}
J.L. developed the framework,conducted the experiments and wrote the manuscript; H.S. and L.Y. performed the data analysis; S.T. and L.Z. revised the manuscript. L.Z. conceived and supervised the study. All authors reviewed the manuscript.

\section*{Code availability}
The source code of ST2M and all configurations used in this work are available on GitHub at \url{https://github.com/MustBeOne/ST2M}.

\section*{Data availability}

The source crystal structures and reference property data used in this study originate from the publicly available Materials Project database at \url{https://materialsproject.org}. Model training and evaluation use the MP20 dataset derived from this resource.

\section*{Competing Interests}
The authors declare no competing interests.

\bibliography{manuscript}

\clearpage
\includepdf[pages=-,pagecommand={},fitpaper=true]{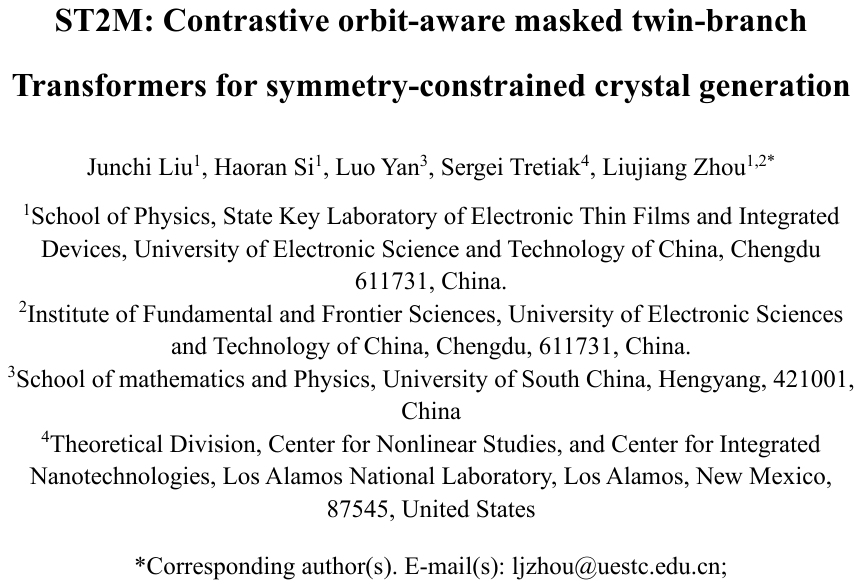}

\end{document}